\documentclass[aps,prd,showpacs,scriptaddress,twocolumn,nofootinbib]{revtex4}

\usepackage{amsmath,amssymb}
\usepackage[normalem]{ulem}
\usepackage{bm}
\usepackage{comment}
\usepackage{xcolor}
\usepackage{soul}

\usepackage[percent]{overpic}
\usepackage{overpic}
\usepackage{natbib}
\usepackage[tiny]{titlesec}
\usepackage{mathrsfs} 

\def\app#1#2{%
  \mathrel{%
    \setbox0=\hbox{$#1\sim$}%
    \setbox2=\hbox{%
      \rlap{\hbox{$#1\propto$}}%
      \lower1.1\ht0\box0%
    }%
    \raise0.25\ht2\box2%
  }%
}

\newcommand{\bea}{\begin{eqnarray}}
\newcommand{\eea}{\end{eqnarray}}
\newcommand{\bma}{\begin{pmatrix}}
\newcommand{\ema}{\end{pmatrix}}
\newcommand{\equalhat}{\mathrel{\hat{=}}}

\usepackage[T1]{fontenc}
\usepackage{ae,aecompl}
\newcommand{\WVU}{\affiliation{Department of Physics, West Virginia University, PO Box 6315, Morgantown,
West Virginia 26506, USA}}
\newcommand{\BNU}{\affiliation{Gravitational Wave and Cosmology Laboratory, Department of Astronomy, Beijing Normal University, Beijing 100875, China}} 

\begin{document} 

\title{Alternative route towards the change of metric signature} 
    \begin{abstract}
Beginning with Hartle and Hawking's no-boundary proposal, it has long been known that the pathology of a big bang singularity can be suppressed if a transition into Riemannian (Euclidean) metric signature (the usual singularity theorems become invalid in this region) occurs when we track back along cosmic time. A vital component of this type of models, that needs to be clarified, is the set of junction conditions at the boundary between the two signature regimes. In the traditional approach, the signature change occurs in the temporal sector through a switch of sign in the lapse-squared function. Motivated by more straightforward connections with the big bang cosmology, we explore here an alternative whereby the spatial metric eigenvalues change sign instead, so that the Riemannian side is purely timelike. We investigate the junction conditions required in this case. 
    \end{abstract}

\author{Fan Zhang} \BNU \WVU

\maketitle

\section{Introduction and motivation}\label{sec:Intro}
\subsection{The standard approaches}\label{sec:Intro1}

Discussions on signature changing spacetimes were arguably ignited by Hartle and Hawking's no-boundary proposal for the initial conditions of the universe \citep{1983PhRvD..28.2960H,1984NuPhB.239..257H}. The study of the semi-classical approximations \citep{1990PhRvD..42.2458G} to the wave function of the universe, especially the dominating real tunnelling solutions (a real Riemannian spacetime joined onto a real Lorentzian one, with the Riemannian part determining the weighting in the path integral) \citep{1983PhRvD..27.2848V,1990PhRvD..42.2458G}, had garnered some interest. 

While studying the tunnelling solutions, it immediately became clear that it is impossible to transition a solution of the Einstein's equations into the Riemannian signature in an uneventful manner, because even a continuous metric will necessarily become either degenerate or divergent.
In other words, General Relativity (GR), without any relaxations, is not intrinsically capable of dealing with signature changes (the transition surface is at best a mild singularity). However, if one is only interested in semi-classical approximations to quantum wave functions, the classical Einstein's equations only need to be ``almost'' satisfied, in the sense that some pathologies on the transition surface is allowed so long as they do not spoil the steepest descent considerations by making a divergent contribution to the action \citep{1990PhRvD..42.2458G}. 
Furthermore, even if we throw away such leniency afforded by quantum mechanics, and consider, as in Refs.~\cite{1992CQGra...9.1535E,1992GReGr..24.1047E}, purely classical set-ups, arguments can be made that suitably weaker versions of GR equations are not outrageous, since after all, there are many situations which the standard GR formalism cannot handle, such as when it comes to singularities inside black holes or impulsive gravitational waves, that do not appear to be prohibited by nature. 

Broadly speaking, depending on the functional space from which one draws solutions to the Einstein's equations that are formally ill-defined (not just singular in a differential equation sense like when some higher derivative terms vanish; some quantities appearing in the equations may become divergent and thus not defined) at a change-of-signature boundary, two types of junction conditions have been proposed in literature (both for when a purely spatial Riemannian side is reached via the temporal eigenvalue of the metric switching sign):
\begin{itemize}
	\item *1: A more flexible one (e.g., Refs.~\cite{1990PhRvD..42.2458G,1991GReGr..23..967D,1992CQGra...9.1535E,1992GReGr..24.1047E,1993PhRvD..48.2587D,1996JMP....37.5627D}) allowing for discontinuous metrics with a continuous but not necessarily vanishing extrinsic curvature of the signature change surface $\Pi$, suitable for distributional solutions. The Einstein's equations themselves remain ill-defined at $\Pi$, so by ``the distribution is a solution'', those authors mean that it satisfies the equation at any point away from that surface, while the equation is suspended on $\Pi$. 

\item *2: Or a more restrictive one (e.g., Refs.~\cite{1992CQGra...9.1851H,1993gr.qc.....3034H,1993CQGra..10.2363K,1993CQGra..10.1157K,1994RSPSA.444..297K}) requiring the metric to be continuous and the extrinsic curvature to vanish when computed from both sides. This set of conditions is suitable for smoother solutions satisfying a regularized version of the Einstein's equations that are not suspended on $\Pi$. Specifically, those offending ill-defined quantities are in fact well-defined off of $\Pi$, so their limits can possibly be obtained through a process asymptoting to $\Pi$, and the broken expressions are then replaced by such limits (and strong junction conditions are required for these limits to exist). One must note that only the covariant form of the equations are regularized, and the inverse metric still diverges, so not everything is made regular in this approach.  

Since the extrinsic curvature is the time derivative of the spatial metric (its trace is essentially the rate at which spatial volume grows), its suppression is often said to imply stationarity. Indeed, similar analysis on other fields propagating on the signature changing background also analogously possess vanishing velocities. This is easy to see from a naive limit-taking analysis of a toy massless Klein-Gordon equation 
\begin{align} \label{eq:Toy}
\varphi^{;a}{}_{;a} = \frac{1}{\sqrt{|g|}}\left(\sqrt{|g|}g^{ab} \varphi_{,b}\right)_{,a} =0\,,
\end{align}
where semi-colon denotes covariant derivative and comma denotes partial derivative. The early part of the Latin alphabet will denote spacetime indices, and the middle part the spatial ones. 
Let $g^{ab}$ be diagonalized in our choice of two dimensional (for illustration) coordinates $(t,x)$ into 
\bea
\bma
-\lambda_t(t,x) & 0 \\
0 & \lambda_x(t,x)
\ema\,,
\eea
then the equation becomes 
\begin{align} \label{eq:KGFull}
&\frac{\lambda_x(\lambda_{t,x})}{\lambda_t}\varphi_{,x} + \lambda_{t,t}\varphi_{,t} + \lambda_t\left( 2 \varphi_{,tt}- \frac{\lambda_{x,t}}{\lambda_x}\varphi_{,t}\right)\notag \\
& - \lambda_{x,x}\varphi_{,x}-2\lambda_x\varphi_{,xx}=0\,.
\end{align}
When approaching the temporal signature change surface $\Pi$, we must have $\lambda_t \rightarrow \infty$ (since it is an entry in the inverse metric) and generically also $\lambda_{t,t} \rightarrow \infty$ at an even faster pace, 
resulting in the requirements of $\varphi_{,t}\rightarrow 0$ and $\varphi_{,tt} \rightarrow 0$ in order for the equation to admit a well-defined limit on $\Pi$.

\end{itemize}
There are essentially two steps involved in deriving these conditions. First is to evoke more or less the generic Darmois junction condition (denoted $\mathcal{C}_g$ below) that the surface metric implied (through pullbacks of the embedding maps) by either side must agree so there is a well-defined three-geometry for the boundary surface, and also that the extrinsic curvatures computed on either side must agree to avoid having to confine a stress-energy tensor onto the spacelike boundary (matter worldlines cannot be entirely confined to a spacelike surface) \citep{Darmois1927,Israel:1966rt,MTW}. Although these conditions are derived in the constant signature case, they essentially remain unchanged in the signature-changing situation (note that with *1, the jump is in the time-time component of the metric, while the spatial part remains continuous, so the implied intrinsic spatial geometries from the two sides still agree). 

The second type of requirements (denoted $\mathcal{C}_s$ below) is specific to the singular (with degenerate or discontinuous metric) signature-changing situation. With *2, $\mathcal{C}_s$ is the vanishing of the matching extrinsic curvatures, which allows a version of the Einstein's equations to be imposed on the transition surface, but is unsurprisingly quite rigid \cite{1994CQGra..11L..87H}.  The *1 approach on the other hand aims for more flexibility by not imposing any $\mathcal{C}_s$ at all, arguing that the extra step of regularizing a singular equation is more a matter of choice than necessity \cite{2001GReGr..33.1041D}. The price it pays is a relaxation of the sense in which the resulting solutions are unique \cite{1992CQGra...9.1535E,1994CQGra..11L..87H}. The differences between the approaches reflect alternative philosophies, perhaps of how universally valid the standard form of the Einstein's equations should remain when its usual underlying assumptions are tempered with. 

\subsection{An alternative} \label{sec:Intro2}
In this paper, we investigate an alternative mechanism by which a signature change can be achieved, following more closely the approach of *2, since we wish to see if the restrictions imposed by the regularization procedure, onto the initial conditions (for our Lorentzian universe) lied down on our transition surface $\Sigma$, can help explain some cosmological fine-tuning issues. So the equations of motion of metric and matter, for which the initial conditions are meant, must not be suspended on $\Sigma$. 

We begin by noting that while having the temporal metric eigenvalue ($1/\lambda_t$ in the notation of Eq.~\ref{eq:KGFull}, since $\lambda_t$ is an eigenvalue of the inverse metric) going through zero (we shall call this approach route A in this paper), either continuously or with a jump, is taken to be the default in previous literature, it is not the only way for the metric signature to change. Having it going through $\infty$ (equivalently $\lambda_t$ through zero) is also valid, since $\infty$ is just the antipodal end of the stereographic projection circle of the real line. 
However, with this approach (route A') in its raw form, the integration measure $\sqrt{-g}$ diverges on $\Sigma$, which has adverse side effects with quantum path integrals (the logic of steepest descent that makes our classical investigation useful in a quantum context may be spoiled \citep{1990PhRvD..42.2458G}). A related approach that removes this problem is to have $1/\lambda_x$ go through zero instead, so that the signature becomes Riemannian not because time changes sign, but because the spatial signature reverses. This alternative (route B) is related to route A' since the $1/\tilde{\lambda}_t = \lambda_x/\lambda_t$ of the conformally rescaled metric $\tilde{g}_{ab} = g_{ab} \lambda_x$ (that shares the same causal structure as $g_{ab}$, such as those depicted in the figures below) goes through $\infty$. I.e., when the physical metric transitions via route B, the conformal metric changes via route A'. 

\begin{figure}[t]
  \centering
\begin{overpic}[width=0.89\columnwidth]{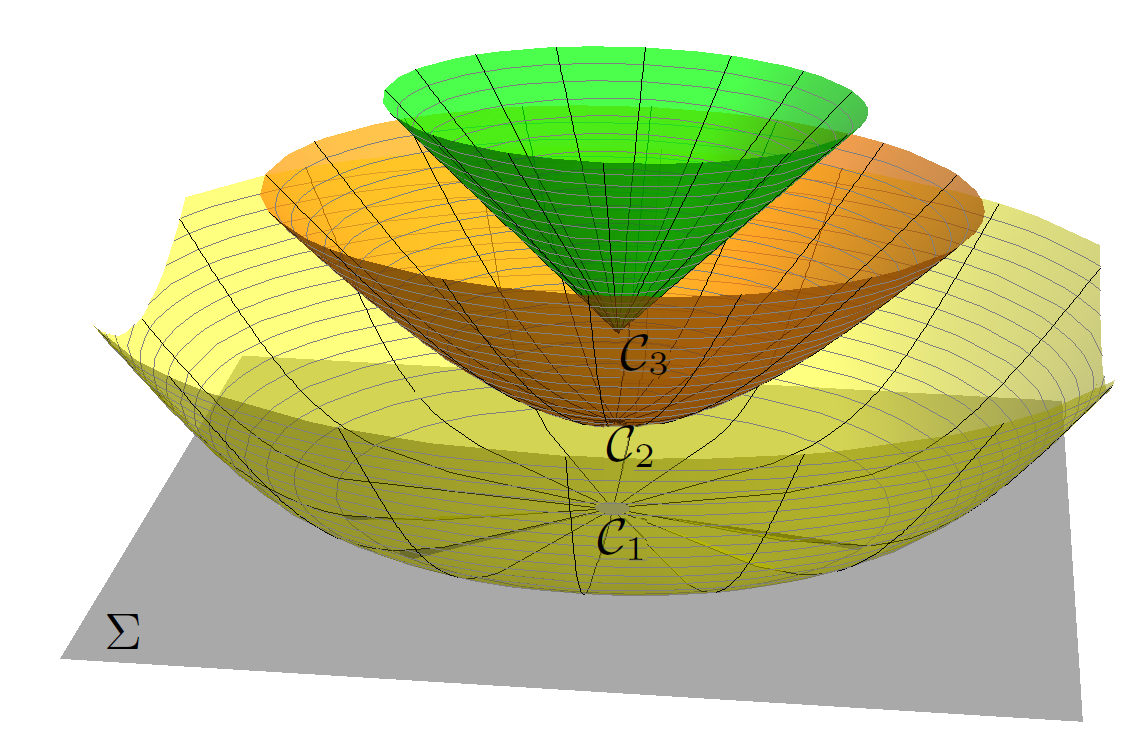}
\end{overpic}
  \caption{As the base points $\mathcal{C}_3$, $\mathcal{C}_2$ and $\mathcal{C}_1$ incrementally approach the signature change surface $\Sigma$, their null cones flatten out (and return to rising more steeply once they are sufficiently far away from $\Sigma$).}
	\label{fig:ConeStack}
\end{figure}

This route B had not been examined in any of the previous literature that we found, and investigating it is the subject of this paper. 
Aside from filling in a gap in literature to achieve pedagogical thoroughness, we note that route B possesses some features that might help make it physically relevant:
\begin{enumerate}
\item 
As compared to route A, it is more straightforward to make connections with our actual universe when we adopt route B, because 
the Friedman-Lema\^{i}tre-Robertson-Walker (FLRW) metric 
\begin{align} \label{eq:ConfTrans}
\quad \quad g_{ab}dx^adx^b =& a(t)^2\tilde{g}_{ab}dx^adx^b\notag \\
=& -dt^2 + a(t)^2 \hat{\gamma}_{ij}dx^i dx^j \notag \\
\equiv & -dt^2 + a(t)^2\left(d\chi^2 + \zeta^2_{\kappa}(\chi)d\Omega^2\right)\,, 
\end{align}
whereby
\bea
d\Omega^2 = d\theta^2 + \sin^2\theta d\phi^2 \,, 
\eea
and 
\bea
\zeta_{\kappa} = \left\{
\begin{array}{ll}
        \sin\chi, & \text{for } \kappa =1\\
        \chi, & \text{for } \kappa =0\\
        \sinh\chi, & \text{for } \kappa =-1\,,
\end{array}\right. 
\eea
is automatically a route-B-compatible metric. 
This means that, within route B, a transition into a Riemannian region can occur at the beginning (where $a(t)=0$ so the spatial metric becomes fully degenerate and ready to be continued further into negative definiteness) of the prevailing cosmological model, extending it beyond the big bang\footnote{Note that contrary to common pictorial depictions, the big bang is not necessarily a single point, just a co-dimension one surface with a degenerate intrinsic metric -- much like how distances along a null ray vanishes, yet the null ray is not a single point. See Sec.~\ref{sec:LightLike} below for more details. }, but without needing significant alterations to the currently prescribed post-big-bang evolution, which wouldn't have been economic since any such alterations must be re-reconciled with observations. 

We will keep the subsequent discussion in this paper general and not specialize to FLRW unless specifically noted. Nevertheless, it is helpful to always have this particularly well-studied and physically relevant special case in mind for intuition building.

\item Route B corresponds to the light cones opening up as one approaches the change of signature surface $\Sigma$ from the Lorentzian side (see Fig.~\ref{fig:ConeStack} for a visual depiction), since equal temporal increments would require increasingly greater spatial coordinate intervals to compensate in the $g_{ab}dx^adx^b =0$ equation for the null rays. 
As suggested by Fig.~\ref{fig:ConeStack} and will be discussed in more details in Sec.~\ref{sec:Morphology} below, the null cone structure is removed (it cannot exist in the Riemannian side) in route B via the future and past null cones opening up to collide and annihilate each other, so it is the spacelike region that is squeezed out of existence, and the Riemannian side is purely temporal, as the metric signature obviously confirm. In contrast, the cones disappear in route A (*2) by separately closing up into a couple of half lines which then vanish beyond $\Pi$. In other words, the timelike regions are the ones taken out in that approach and the Riemannian side is purely spatial. 

A complication of that latter method is then that it takes constructive efforts (e.g., use different definitions for the geodesic Lagrangian when in alternative signature regimes \cite{1992GReGr..24.1047E}) to make timelike geodesics thread through $\Pi$, since if left alone, they would have disappeared together with the timelike regions. In contrast, such intervention is unnecessary with route B, whose Riemannian side is capable of hosting timelike curves. That such a continuation of timelike geodesics is required in the first place is due to the desire to show that the signature change scenario no longer suffers geodesic incompleteness, so that the big bang singularity is indeed removed in that particular sense, and one stays faithful to the original no-boundary proposal of \cite{1983PhRvD..28.2960H}. This amelioration is possible because the usual singularity theorem \cite{1965PhRvL..14...57P,1966PhDT.......101H} needs some causal properties that are no longer available in the Riemannian regime \cite{1992CQGra...9.1535E}. 

\item 
Following a procedure closely mimicking that of *2 but for route B, we obtain once again strong $\mathcal{C}_s$ conditions, but now including an additional one ($\mathcal{C}^2_s$ of Sec.~\ref{sec:CondEin}) enforcing the vanishing of spatial derivatives on $\Sigma$, in addition to the temporal stationarity. Furthermore, the lapse function within route B can be set to a constant, so that even more components of the four-metric's derivatives vanish as compared to route A.   
Because these metric derivatives contribute to the curvature tensors, their suppression is beneficial for realizing the uniformity condition on the big bang, that's envisaged by the Weyl curvature conjecture \cite{1989NYASA.571..249P}\footnote{Incidentally, it was noted in this paper that something along the lines of the Hartle-Hawking no-boundary proposal may lead to the required condition.} to start the universe off on low entropy (see also \cite{2014arXiv1406.3057C}). A signature change universe via route B thus offers up an intriguing new way to supplement inflation in its quest to solve some cosmic puzzles.

\end{enumerate}

In the rest of the paper, we turn to the details, beginning by establishing some basic properties of a route B transition in Sec.~\ref{sec:Conformal}, before finding the junction conditions in Sec.~\ref{sec:Junction}. We finally conclude in Sec.~\ref{sec:Conclusion} with a discussion on the many studies required to more thoroughly explore the viability and properties of route B. For the Lorentzian side, we adopt signature $(-,+,+,+)$ and the Riemannian side subsequently has $(-,-,-,-)$.

\section{Large scale features} \label{sec:Conformal}

\subsection{The signature morphology} \label{sec:Morphology}
\begin{figure}[t]
  \centering
\begin{overpic}[width=0.99\columnwidth]{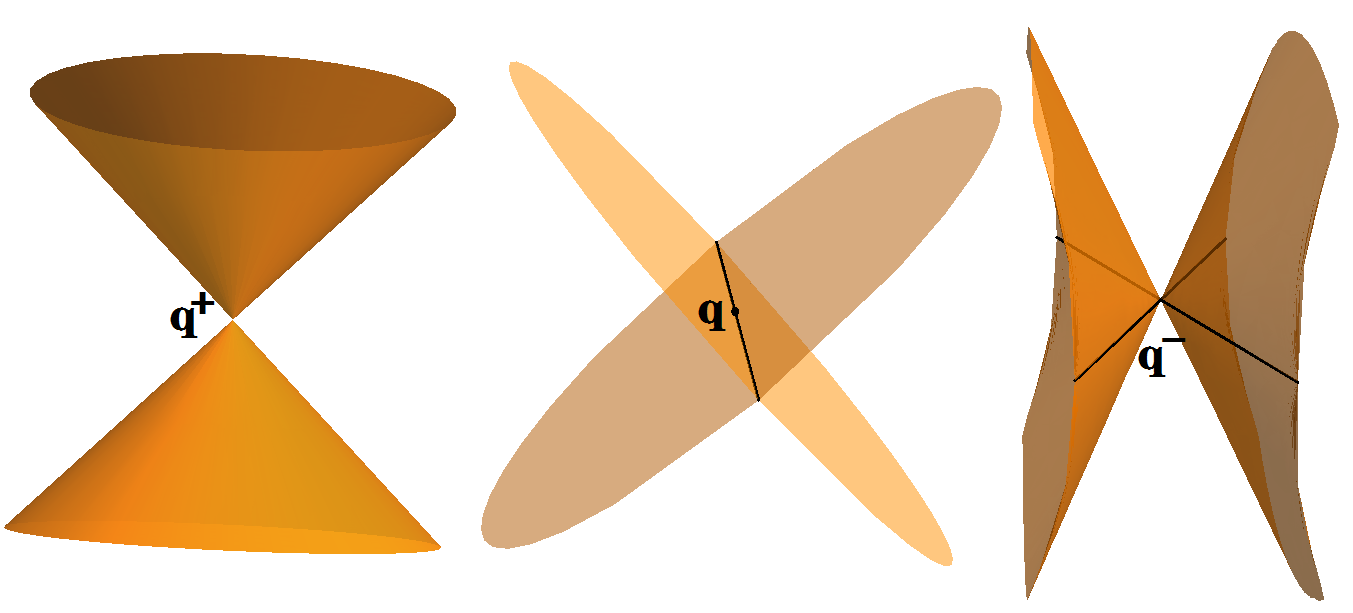}
\end{overpic}
  \caption{If a single eigenvalue of $\gamma_{ij}$ turns negative at $q$, then at a later time $q^+$ along a timelike path through $q$, the future and past null cones remain separated as per usual with a Lorentzian metric signature; at $q$, the two very elongated cones touch along a (black in the figure) line that's the eigenvector (eigenvectors are not directional, so the black line is a full and not half line) corresponding to the eigenvalue that now becomes zero, and thus vectors along this direction now have zero norms; at some earlier time $q^-$, the two cones partially ``annihilate'', splitting the black line. The timelike interior of the cones join up through the opened up gap between the pair of black lines indicating the intersection lines between the cones. When all three eigenvalues change sign at the same point $q$ (not plotted), the cones will open up towards each other as we approach $q$ from $q^+$, and the entireties of the two cones collide and annihilate at $q$, so there are no null cones at $q^-$ and all directions are timelike -- the spacetime becomes Riemannian. }
	\label{fig:Merge}
\end{figure}

A complication that route A did not suffer but route B must now face is that since three dimensions now switch signature, there is the possibility that the three switches occur sequentially, instead of simultaneously as in the FLRW example. Specifically, consider the generic metric in the 3+1 form \cite{2008GReGr..40.1997A}
\begin{align} \label{eq:Synch}
&g_{ab}dx^a dx^b \notag \\
&\quad =
-\alpha^2 dt^2 + \gamma_{ij}\left( \beta^i dt + dx^i \right)\left( \beta^j dt + dx^j \right)\,,
\end{align}
where we fix the gauge freedoms by setting lapse $\alpha \equiv 1$ and shift $\beta^i \equiv 0$ so as to pick Gaussian normal (synchronous) coordinates, whose temporal coordinate curves are timelike geodesics. Starting from an arbitrary coordinate system $x^{\bar{a}}$, we can find the Gaussian normal coordinates by solving the Hamilton-Jacobi equation 
\bea
g^{\bar{a}\bar{b}}S_{,\bar{a}}S_{,\bar{b}}=-1\,,
\eea
for which real solution $S$ (it is to be the new time coordinate) exists even as signature is allowed to vary, because we always have at least one timelike dimension within route B. A subtlety is that at places of signature change, some spatial\footnote{We will slightly abuse terminology in the interest of brevity and assign the label ``spatial'' to the other coordinates that are not $t$, even though their associated dimension can become null or timelike.} components in the inverse metric $g^{\bar{a}\bar{b}}$ may diverge, but well-defined limits exist for these locations if the spatial derivatives of $S$ simply vanish sufficiently quickly there. Once the solution is found, the Gaussian normal coordinate system can be constructed by following the standard textbook recipe. In this new coordinate system, that we adopt for expositional clarity, $\gamma_{ij}$ is positive definite in a usual Lorentzian region, but its eigenvalues can transition, either one at a time or several together, into negative values. We can understand what this physically means by examining what happens to the null cones when one or more eigenvalues turns to zero and then negative through a transition point $q$. The situation is depicted in Fig.~\ref{fig:Merge}, and the local tangent space geometry can be intuited as future and past null cones colliding and ``annihilating'', allowing their timelike interiors to merge. 

The possibilities of partial sign switches and thus more diversified signature configurations are intriguing, but physically problematic. For example, a scalar field in a signature $(-,-,+,+)$ spacetime region would propagate via an ultrahyperbolic equation, which is generically (unless nonlocal constrains are imposed \citep{2009RSPSA.465.3023C}) ill-posed  \citep{1953mmp..book.....C} when evolved off of any Cauchy surface (on the other hand, an elliptic equation in an Riemannian region admits well-posed boundary value problems).  
Beyond the scalar field, Ref.~\cite{2001PhLB..520..159V} also showed that fields with finite spins greater than zero cannot be defined in a signature $(-,-,+,+)$ spacetime region (they are however allowed in a Riemannian region). This implies a rather strange requirement where a Dirac field describing say, electrons, can exist in the Lorentzian region, but must somehow collude with spacetime in a fashion far beyond simply warping it, and stop existing (not just becoming zero in amplitude) as soon as one spatial direction mutates.
 
We therefore need at least two spatial directions to switch simultaneously, leading to a time-space swapped Lorentzian spacetime of signature $(-,-,-,+)$. All massive particles must now become essentially long-lived ``tachyons'' \citep{1997CQGra..14L..69T}, since they now move outside of lightcones (centred on the remaining spatial direction) in order to follow timelike worldlines. Because the mathematics for a quantum field theory in this region is the same as in the regular Lorentzian signature (in fact, the sign convention adopted in particle physics is the opposite of that used by relativists, and this paper, so no sign changes are even needed when lifting formulae from books), one are then faced with all the vacuum instability issues and other pathologies that tachyons bring. 
We therefore assume, from here onwards, that sequential sign switches are forbidden, and that all three spatial dimensions switch simultaneously, giving us $(-,-,-,-)$ straight away. 

\subsection{The FLRW junction surface}\label{sec:LightLike}
Continuity of the metric requires that any curve linking two points of different signatures must intersect the bounding wall $\Sigma$ at least once, so $\Sigma$ should at most have codimension one (the curve itself takes up one codimension, and if there exists another, the curve's intersection point with $\Sigma$, as well as the surrounding sections to preserve continuity, can be shifted in that direction to avoid $\Sigma$), but does not need to be a constant $t$ surface. The case of the highly symmetric FLRW is much simpler though, and due to its cosmological relevance, worthy of us taking a little detour to clarify. We emphasize though that most of our discussions on the junction conditions in Sec.~\ref{sec:Junction} are not confined to this case, and are valid for generic $\Sigma$s. They are local considerations relating to limit-taking procedures along an arbitrary single timelike curve threading through $\Sigma$ at a single point, and are as such independent of the larger scale properties of $\Sigma$.   

The first thing one notices is that the FLRW $\Sigma$ is null, since $g_{ab}dx^adx^b$ vanishes at $a=0$ for any separations confined to $\Sigma$ (with $dt=0$ since $a$ is a function of $t$ only). It should be noted though, points on $\Sigma$ can be macroscopically separated yet null-related just like two points along a null ray, in which case $\Sigma$ is not a single point as often depicted for the big bang, just like a null ray is not a single point. This situation arises because the Lorentzian metric is rather pathological for the purpose of defining open sets (metric balls are noncompact) and studying topology (thus the frequent adoption of a positive definite auxiliary metric in some topological studies, see e.g., \cite{Beemfest,1979grec.conf..212G}). Baring any direct observational consequences of the Riemannian side from which the topology of $\Sigma$ may become more obvious, the best hope we have to ascertain its nature may be to assume global hyperbolicity of the Lorentzian side, whose topology should then be a direct product $\mathbb{R} \times \Sigma$ \cite{1990PhRvD..42.2458G}. Since our universe is not one dimensional, $\Sigma$ cannot be a zero dimensional point. 

More explicitly, the observational evidence is that the spatial slices of our universe appear to be flat \cite{2018arXiv180706209P}, and thus could well be infinite in extent\footnote{They could also be flat tori or other twisted alterations \cite{Riazuelo:2003ud}, but a point would still have the wrong dimension for a boundary of the 4-D Lorentzian universe, and would instead be an interior point, resulting in the big bang cosmology already being ``no-boundary'' even without introducing a Riemannian region, depriving us of a place to prescribe initial conditions (necessary for Cauchy evolution on the Lorentzian side) on.}. It would then be quite strange for such an infinite noncompact plane to instantaneously collapse into a single point (a compact singleton) as soon as the scale factor reaches precisely zero, when it would still be noncompact for any infinitesimal value of $a>0$. In that scenario, the early universe would not resemble the collar neighbourhood of $\Sigma$, which would obviously adversely affect our ability to evolve initial conditions off of $\Sigma$ to uniquely determine the Lorentzian side of the universe. 

Incidentally, in the case of the flat slicing of de Sitter (see e.g., Fig.~1 of Ref.~\cite{2002PhRvD..65h3507A}) serving as an isometry to an inflationary FLRW, the finite comoving observers (those labelled by finite spatial comoving coordinates) do in fact all get packed into a single asymptotic point of the de Sitter spacetime when traced back in time. The abrupt jump issue in this case is resolved by pulling in points from comoving spatial infinity to form an extended noncompact border surface. This is fine for de Sitter, since points on this surface are just regular points inside the actual de Sitter spacetime. Their carrying infinite spatial coordinates is simply due to the flat foliation coordinates being singular (a symptom is that this coordinate system cannot be extended beyond this border to cover the other half of de Sitter). However, for the actual FLRW universe, 
there is no reason to believe that the comoving coordinate system, as preferred by the observed motion of matter, is ill-chosen and ill-behaved, so similar infinity points would likely genuinely reside on the spatial compactification boundary. In other words, they are outside of the actual spacetime (similar to how the future null infinity $\mathscr{I}^+$ \cite{1965RSPSA.284..159P} is outside of an asymptotically flat spacetime itself), and are mathematical constructs not in fact physically available, to smooth out the jump, or to prescribe junction conditions on. In short, while the inflationary FLRW and de Sitter are isometric for the post-big-bang segment, they likely differ when it comes to the topological structure of the big bang itself, which is not a radical prospect given that they already differ on what lies beyond. 

Finally, as an aside, it is also worthwhile noting that the FLRW big bang is sometimes said to be spacelike, but this characterization is under the conformal metric rather than the physical metric, and the choice is not unique. Specifically, there is a well established field of study on the ``conformal gauge singularities'' (regarding the big bang singularity as being due to the special ``conformal gauge choice'' in a conformal class of mostly regular metrics) \citep{1985CQGra...2...99G,1999AnPhy.276..257A,2008AnPhy.323.2905L,2009JMP....50k2501L,2010JPhCS.229a2013T}. Even the well-posedness of the Cauchy problem for various matter content types have been proven for this construct \citep{1993RSPSA.443..473N,1999AnPhy.276..257A,1999AnPhy.276..294A,2000AnPhy.282..395A,2003CQGra..20..521T,2007CQGra..24.2415T}. It is also useful for us to think about the causal structure of the spacetime using the conformal metric $\tilde{g}_{ab}$, but we stop short of carrying out the additional temporal transformation $t\rightarrow \tau$ defined by 
\bea \label{eq:ConformalTime}
\frac{d\tau}{dt}=\frac{1}{a(t)} \,.  
\eea
After this extra layer of coordinate transformation, the FLRW metric becomes conformally flat. However Eq.~\eqref{eq:ConformalTime} is singular at $a=0$, and since $t$ is the intrinsic clock carried by physical comoving observers, 
results obtained under $\tau$ must be fed through an additional singular transformation before it can be translated back into predictions on physical experimental outputs. The reward for this extra trouble is flexibility. Specifically, since $dt/d\tau =0$ at the big bang, the condition of moving along the constant $t$ surface, as expressed by $dt=0$, can be satisfied by any finite $d\tau$ choice. Instinctively, one picks $d\tau=0$ which gives a conformally spacelike (under the physical metric it is still null) big bang, but one could actually equally well choose other $d\tau$ that makes it conformally null or even timelike. The arbitrariness is because that essentially, via an infinite stretching, the zero-thickness three dimensional $\Sigma$ got stretched into a four dimensional object. While people still customarily pick out a 3-D surface in there and call it the big bang, it perhaps should have been the whole 4-D totality. Regardless, if one holds the view that this newly inserted internal structure to the big bang is physical, then its flexibility would allow for establishing beautiful mathematical infrastructures. We will remain more parsimonious in this paper though, and formulate the junction conditions under the physical metric. 

\section{The junction conditions} \label{sec:Junction}
A singular differential equation can be well-defined at its singular set, e.g., $x \beta_{,xx}+ \beta=0$ at $x=0$. However, with our toy Eq.~\eqref{eq:KGFull} or Einstein's equations, the coefficient functions appearing in the equation or the curvature expressions become divergent or otherwise ill-defined (e.g., $0/0$) on $\Sigma$. So strictly speaking, the equations are not merely singular; they are not formally defined there. Nevertheless, we could follow the approach of \cite{1993CQGra..10.2363K} and regularize the offending divergences by imposing strict junction conditions, so that the equations admit well-defined limits on $\Sigma$. Solutions satisfying such conditions can then be sought such that the $\Sigma$ limits of the left and right hand sides of the equations match. This distils a set of equations of motion to be satisfied on $\Sigma$, so physics won't be left completely arbitrary there, but as already alluded to in Sec.~\ref{sec:Intro1}, $\Sigma$ cannot be rendered completely regular. At the very least, the inverse metric still diverges, and some of the Carminati-McLenaghan curvature invariants \citep{Carminati:1991} might do so as well. Our present endeavour is a modest attempt at partially resolving the big bang singularity in order to glean some information on the likely behaviour of the important classical saddle point solutions; it is not aimed at removing the singularity altogether, a task for which an understanding of quantum gravity is probably required (but the intriguing possibility of accomplishing it even at a classical level, perhaps through the adoption of more topology-friendly auxiliary metrics, should not be dismissed out of hand; for such an investigation, our study would serve as a first step to demonstrate how far one can go without bringing in additional infrastructures, and to identify the remaining problems they must solve, thereby clue us in on where new physics/mathematics might come in, as well as what they might look like). 

\subsection{The Einstein's equations} \label{sec:CondEin}
\subsubsection{The method} \label{sec:EinMethod}
We begin with the left hand side of the Einstein's equations. 
Following standard literature \cite{MTW}, under Gaussian normal coordinates, the Einstein tensor can be written in the $3+1$ form as
\begin{align}
G_{tt}=& \frac{1}{2}{}^{(3)}R + \frac{1}{2} \left[ K^2 - \text{tr}(K\cdot K)\right] \,, \label{eq:G1}\\
G_{ti} =& -K_i{}^m{}_{|m}+K_{|i} \,, \label{eq:G2} \\
G_{ij} =& {}^{(3)}G_{ij} - \bigg\{ (K_{ij}-K \gamma_{ij} )_{,t} 
+ 2 K_{ik}K^{k}{}_j 
- 3K K_{ij} 
\notag \\
&+ \frac{1}{2}K^2\gamma_{ij}+ \frac{1}{2} \text{tr}(K\cdot K)\gamma_{ij}
\bigg\}\,,\label{eq:G3} 
\end{align}
where $K_{ij}=- \gamma_{ij,t}/2$ is the extrinsic curvature of the constant $t$ slice (not necessarily coincident with $\Sigma$), and the vertical bar denotes 3-D covariant derivative. 
The first two equations do not contain temporal derivatives, and are the Hamiltonian and momentum constrains respectively. The third equation tells us how to evolve the metric in time. 
Note that these expressions are valid on both sides of $\Sigma$, since unlike with route A, there is no change to the norm of the normal vector $\partial_t$ of the spatial slices within route B, thus none of the explicit signs in Eqs.~\eqref{eq:G1}-\eqref{eq:G3} needs to change; the signature changes are all hidden inside the symbolic spatial quantities, just as they are all hidden inside the $\gamma_{ij}$ in the metric (see Eq.~\ref{eq:Synch}). We therefore will not explicitly distinguish between the Lorentzian and Riemannian sides in the derivations below, since all expressions are identical. 

Because 
$\gamma^{ij}$ is the source of divergences at $\Sigma$, the terms 
\begin{align}
K \equiv \gamma^{ij}K_{ij}\,,
\quad
\text{tr}(K\cdot K) \equiv \gamma^{ij}\gamma^{kl}K_{ik}K_{jl}\,, 
\notag
\end{align}
as well as $K_i{}^{m}{}_{|m}$, $K_{ik}K^{k}{}_j$, 
and since (also due to other contractions with the inverse metric in trace-taking computations)
\bea \label{eq:Connection}
{}^{(3)}\Gamma ^i_{{jk}}=\frac{1}{2} \gamma^{{il}} \left(\gamma_{{kl,j}}-\gamma_{{jk,l}}+\gamma_{{lj,k}}\right) \,,
\eea
also ${}^{(3)}R$ and ${}^{(3)}G_{ij}$, could all diverge there. The goal, following the arguments of Ref.~\cite{1993CQGra..10.2363K}, is to see what conditions arise from demanding that $G_{ab}$ remains bounded in the $\Sigma$ limit. It should be noted that this approach demands component-wise regularity for the Einstein tensor (because the Einstein's equations are in component form), but being explicit tensor components, the expressions \eqref{eq:G1}-\eqref{eq:G3} depend on the underlying coordinate basis onto which the tensor is decomposed, and this basis could be ill-behaved even when the underlying geometry is perfectly fine (e.g., if caustics develop for the congruence of timelike geodesics underlying the Gaussian normal coordinate system, due to a bad choice of initial velocities). This coordinate singularity issue is familiar and not specific to the problem at hand, but it is nevertheless worth emphasizing that it implies the conditions $\mathcal{C}^{1/2}_s$ we obtain in the next section are (unfortunately unavoidably) sufficient but not necessary. 

They are also quite strong in another way, as they will demand that all the terms in $G_{ab}$ that could possibly diverge would instead remain finite, in an individual term-wise fashion. There is of course also the possibility that divergences cancel across terms. To find these cases, one needs to solve differential regularity equations derived from the condition that the divergent terms in Eqs.~\eqref{eq:G1}-\eqref{eq:G3} are curbed, which is technically difficult without assuming symmetries to simplify expressions, but doing so would defeat the purpose of trying to find out what kind of constraints that regularity at $\Sigma$ would place on our universe. Instead, we deploy generic considerations to argue that such solutions would unlikely be numerous (or indeed exist at all), so at the very least, the solutions given by $\mathcal{C}^{1/2}_s$ would not be unlikely as physically relevant junction conditions from a statistical point of view. 

We begin by noting that Eqs.~\eqref{eq:G2} and \eqref{eq:G3} contain terms involving both one and two factors of the inverse spatial metric $\gamma^{ij}$, which diverge at different rates and have to be treated separately. Schematically, write $\gamma^{ij} \sim 1/\zeta$ with $\zeta \rightarrow 0$ when approaching $\Sigma$, then Eq.~\eqref{eq:G2} or \eqref{eq:G3} could be stylized as 
\bea \label{eq:RawCancel}
\frac{A(\zeta)}{\zeta}+\frac{B(\zeta)}{\zeta^2}=\frac{1}{\zeta}\left(A+\frac{B}{\zeta}\right)\,,
\eea
where $A$ and $B$ are non-divergent at $\zeta=0$ since we have collected all the problematic terms into powers of $1/\zeta$. Regularity then requires that 
\bea \label{eq:Reg1}
A = -\frac{B}{\zeta} + \mathcal{O}(\zeta)\,,
\eea
and since $A$ can not diverge (but can be nonvanishing) when $\zeta\rightarrow 0$, we also need 
\bea \label{eq:Reg2}
B = \mathcal{O}(\zeta)\,. 
\eea
The cross cancellations thus allow for more relaxed $A$ and $B$ than what term-wise regularity would demand, which is $A = \mathcal{O}(\zeta)$ and $B=\mathcal{O}(\zeta^2)$ (these select a subset of solutions to Eqs.~\ref{eq:Reg1} and \ref{eq:Reg2}, and are not alternatives to them). 

The catch is that for each original Eq.~\eqref{eq:RawCancel}, we end up with twice as many regularity conditions \eqref{eq:Reg1} and \eqref{eq:Reg2}. This means that Eqs.~\eqref{eq:G2} and \eqref{eq:G3} would demand $2\times 3$ and $2\times 6$ regularity equations respectively, while Eq.~\ref{eq:G1} adds another few. 
Furthermore, the three eigenvalues $\gamma_{\iota}^e$ could all vanish at different rates, so instead of just two powers as in our stylized example Eq.~\eqref{eq:RawCancel}, there are in fact more distinct divergence rates, spawning a great many regularity equations. On the other hand, there are only 12 independent components in the variables $\gamma_{ij}$ and $K_{ij}$\footnote{We are here taking the Hamiltonian approach of Arnowitt-Deser-Misner (ADM) \cite{2008GReGr..40.1997A}, where $\gamma_{ij}$ and $K_{ij}$ are regarded as independent variables, each marching forward according to a first-derivative-in-time evolution equation. One can of course also take the Lagrangian view and see $\gamma_{ij}$ as the only fundamental variable, governed by a second-derivative-in-time evolution equation. By definition, the regularity equations are there to limit what initial conditions one can place on $\Sigma$ (they are allowed to be under-determining), and for these initial conditions, one can either lay down $6$ initial values for $\gamma_{ij}$ and $K_{ij}$ each, or $12$ for $\gamma_{ij}$ alone (since it is then governed by a second order equation, one should give both Dirichlet and Neumann conditions). The number of required initial values is always $12$, and they must satisfy the $\Sigma$ limit of the regularity equations, which is generically impossible if there are more than $12$ such equations.}, so the coupled set of regularity equations is heavily overdetermined, thus generically does not admit solutions beyond the trivial ones identified by $\mathcal{C}^{1/2}_s$. By triviality\footnote{The regularity equations are stronger than just implicit initial conditions confined to $\Sigma$, since they also constrain the variables at small but nonvanishing $\zeta$ values (for well-defined limits to exist for $G_{ab}$, just having the coefficients to the various powers of $1/\zeta$ vanishing on $\Sigma$ is not enough, they also need to vanish sufficiently quickly as $\Sigma$ is approached). Yet they are more relaxed than the usual exact equations in the interior, since small errors are allowed (e.g., stray $\zeta^2$ terms are allowed for Eq.~\ref{eq:Reg1} due to the $\mathcal{O}(\zeta)$ provision).
For interior equations, the relevant number of free variables drops to $6$ (the number of physical freedoms in a metric; the $6$ freedoms in $K_{ij}$ are removed by its definition as the time derivative of $\gamma_{ij}$, which gives a set of $6$ constraint equations) from the $12$ as for the initial conditions, exacerbating the over-determinacy. 
Alternatively, one may stay with the boundary view and note that each of the regularity equation is stronger than just one initial condition, since one could expand it into powers of $\zeta$ (as surrogate for $t$) around $\zeta=0$ (i.e., the expansion coefficients are evaluated on $\Sigma$), and observe that coefficients to all powers lower than that inside $\mathcal{O}$ must vanish (there is always at least one such coefficient, that of $\zeta^0$, but there could be more), translating into multiple initial conditions. Thus the severity of over-determinacy is underestimated in the main text, although already sufficient for our purpose. The triviality discussed here is in regard to these relaxed interior equations, and a more familiar notion of trivial solutions to overdetermined exact equation systems can be recovered by confining the discussion to $\Sigma$ itself, as we have also done in this footnote.}, 
we mean that individual terms in each equation are all pushed below the ``error budget'' of that equation (e.g., $\mathcal{O}(\zeta)$ for Eq.~\ref{eq:Reg1}) by $\mathcal{C}^{1/2}_s$, so no strict equalities need to be actually enforced (to precisely balance/cancel out between quantities above the error tolerance threshold), resulting in the over-abundance of equations all being rendered inert, left with no chance to conflict with one another. 
 
For an illustrative example of how the error budget bestows flexibility, take the case $A \propto B^p$ with some fixed $p$ prescribed by physics (i.e., there is only one free variable $B$, and the Eqs.~\ref{eq:Reg1} and \ref{eq:Reg2} are overdetermined), and let $B \propto \zeta^q$ be an ansatz solution whose $q$ is up to us to pick. Then if we take up the more relaxed $q=1$ as allowed by Eq.~\eqref{eq:Reg2}, we would have a chance of balancing Eq.~\eqref{eq:Reg1} only in the fine-tuned case of $p=0$. However, if $q=n\geq 2$ as required by $\mathcal{C}^{1/2}_s$, then any $p\geq 1/n$ is comfortably accommodated. This is because, in the latter case, the $B/\zeta$ term does not rise above the error tolerance $\mathcal{O}(\zeta)$ of the overall equation, thus does not require careful cancellation from $A$ (as a result, $p$ does not have to be of any particular value), which is not freely variable and thus is defective for fulfilling this role. Note that although we have used cancellation across different powers of $\zeta$ for our example, analogous considerations, as well as the triviality discussion of the last paragraph in general, also apply to cancellations between terms contributing to the same power (i.e., $A$, and/or $B$, alone could further subdivide into a small number of contributors), which $\mathcal{C}^{1/2}_s$ also excludes.

\subsubsection{The conditions}
Near a temporal coordinate geodesic $\xi$ of our Gaussian normal coordinate system $x^a$ that threads through $\Sigma$, we can construct a principal coordinate system $x^{\check{a}}$ under which $\gamma_{\check{i}\check{j}}$ is diagonalized on $\xi$, by first applying an $\text{O}(3)$ transformation within the spatial tangent space to diagonalize $\gamma_{ij}$ there (this is always possible according to the Spectral Theorem since $\gamma_{ij}$ is non-singular real symmetric; we don't need $x^{\check{a}}$ to be unique), and then lay down the spatial coordinates in an open tube surrounding $\xi$ via the exponential map. Note that we do not normalize $\gamma_{\check{i}\check{j}}$ to unity (the coordinate basis $\{\partial_{x^{\check{a}}}\}$ is not orthonormal), so that the Jacobian transforming between the two coordinate systems remains well-behaved along $\xi$, even as we approach $\Sigma$ (always just a block diagonal matrix with a regular orthogonal matrix for the spatial sector, and unity for the temporal sector). 
Along $\xi$ (where the spatial tangent spaces according to $x^{a}$ and $x^{\check{a}}$ coincide), the spatial tensors on the right hand sides of Eqs.~\eqref{eq:G1}-\eqref{eq:G3} can be computed as their counterparts in the $x^{\check{i}}$ coordinate system multiplied for an appropriate number of times by the $O(3)$ spatial Jacobian, which is never divergent nor degenerate (always full-ranked), thus it suffices to examine the divergences under $x^{\check{i}}$ where the algebraic matrix operations reduce to those between the three eigenvalues $\gamma^e_{\iota},\,\iota \in \{1,2,3\}$ shared by $\gamma_{ij}$ and $\gamma_{\check{i}\check{j}}$, which are positive on the Lorentzian side and negative on the Riemannian side. 

Essentially, we have here a Fermi normal coordinate construction \cite{1922RendL..31...21F,1963JMP.....4..735M} with the addition of a rescaling step (on the parallelly transported spatial basis vectors) to recover the eigenvalues, 
thus ensuring that the divergences are not appropriated by the coordinates and are captured by $\gamma^{\check{i}\check{j}}$. Just like the Fermi coordinates, our principal coordinates covers the entire open tube, but the nice properties such as the metric $\gamma_{\check{i}\check{j}}$ being diagonal is only true on the geodesic $\xi$ itself (a $[\xi]$ prefix below signifies expressions valid only on $\xi$). This is fine for us though, since we are studying the limiting behaviours of quantities as we approach $\Sigma$ along $\xi$, so we only ever need to evaluate such quantities on $\xi$. Therefore, in our computations, $\gamma_{\check{i}\check{j}}$, $\gamma^{\check{i}\check{j}}$ and their temporal partial derivatives (measuring changes along $\xi$) to any order (including $K_{\check{i}\check{j}}$ in particular) are diagonal
\begin{align} \label{eq:SpecialSpatialCoord}
\big[\xi\big]:\,\,&\gamma_{\check{i}\check{j}} = \Big[ \text{diag}_{\iota} \gamma^e_{\iota} \Big]_{\check{i}\check{j}}\,, \quad 
\big[\xi\big]:\,\, \gamma^{\check{i}\check{j}} = \left[ \text{diag}_{\iota} \frac{1}{\gamma^e_{\iota}} \right]^{\check{i}\check{j}} \,, \notag \\
\big[\xi\big]:\,\,&K_{\check{i}\check{j}} = \left[ -\frac{1}{2}\text{diag}_{\iota} \gamma^e_{\iota,\check{t}} \right]_{\check{i}\check{j}}\,, \notag \\
\big[\xi\big]:\,\,&K = -\frac{1}{2}\sum_{\iota}\frac{ \gamma^e_{\iota,\check{t}}}{\gamma^e_{\iota}}\,, \notag \\
\big[\xi\big]:\,\,&K_{,\check{t}} = -\frac{1}{2}\sum_{\iota}\left(\frac{\gamma^e_{\iota,\check{t}\check{t}}}{\gamma^e_{\iota}} - \left(\frac{\gamma^e_{\iota,\check{t}}}{\gamma^e_{\iota}}\right)^2\right)\,, \notag \\
\big[\xi\big]:\,\,&\text{tr}(K\cdot K) = \frac{1}{4}\sum_{\iota}\left(\frac{\gamma^e_{\iota,\check{t}}}{\gamma^e_{\iota}}\right)^2\,,\notag \\
\big[\xi\big]:\,\,&K_{\check{i}}{}^{\check{m}} = \left[ -\frac{1}{2}\text{diag}_{\iota} \frac{\gamma^e_{\iota,\check{t}}}{\gamma^e_{\iota}} \right]_{\check{i}}{}^{\check{m}}\,,
\notag \\
\big[\xi\big]:\,\,&K_{\check{i}\check{k}}K^{\check{k}}{}_{\check{j}} = \frac{1}{4} \left[\text{diag}_{\iota}\frac{( \gamma^e_{\iota,\check{t}})^2}{\gamma^e_{\iota}}\right]_{\check{i}\check{j}}\,. 
\end{align}
We then immediately see that the requirement (c.f., Ref.~\cite{1993CQGra..10.2363K} for route A) 

\vspace{3mm}
\noindent $\mathcal{C}^1_s$: \emph{Temporal derivatives of the spatial metric vanish at least as quickly as the spatial metric itself as $\Sigma$ is approached, in the sense that $\gamma^e_{\iota,\check{t}} = \mathcal{O}(\gamma^e_{\iota})\,,\,\,\gamma^e_{\iota,\check{t}\check{t}} = \mathcal{O}(\gamma^e_{\iota})\,,\,\, \forall \iota$}, 
\vspace{3mm}

\noindent 
is necessary and sufficient to ensure that the following terms
\begin{align} \label{eq:timeterms}
&K^2\,, \quad \text{tr}\left(K\cdot K\right)\,, \quad (K_{\check{i}\check{j}}-K \gamma_{\check{i}\check{j}} )_{,\check{t}}\,, \notag \\ 
&K_{\check{i}\check{k}}K^{\check{k}}{}_{\check{j}}\,, \quad K K_{\check{i}\check{j}}
\end{align}
in $G_{\check{a}\check{b}}$ all individually remain bounded. Note that although $K_{\check{i}\check{j}}$ vanishes on $\Sigma$, its trace $K$ does not need to, since $\mathcal{C}^1_s$ allows for $\gamma^e_{\iota,\check{t}} = \Theta(\gamma^e_{\iota})$, i.e., it allows the numerator and denominator in the $K$ expression in Eq.~\eqref{eq:SpecialSpatialCoord} to vanish equally quickly when approaching $\Sigma$, so the limit of the ratio can be finite but nonvanishing. 

The other terms in $G_{\check{a}\check{b}}$ not appearing in Eq.~\eqref{eq:timeterms} involve spatial derivatives. In general, spatial derivatives of even the off-diagonal entries in the tensorial quantities appearing in Eq.~\eqref{eq:SpecialSpatialCoord} do not necessarily vanish, since these quantities can be non-diagonal off $\xi$. Nonetheless, because the principal coordinate system is constructed via the exponential map, we must have a vanishing connection 
\bea \label{eq:ConnectionExplicit}
\big[\xi\big]:\,\,{}^{(3)}\Gamma^{\check{i}}_{\check{j}\check{k}} =0\,, 
\eea
and subsequently
\bea \label{eq:FirstDerivs}
\big[\xi\big]:\,\,\gamma_{\check{i}\check{j},\check{k}}=\gamma_{\check{i}\check{l}}{}^{(3)}\Gamma^{\check{l}}_{\check{j}\check{k}}+\gamma_{\check{j}\check{l}}{}^{(3)}\Gamma^{\check{l}}_{\check{i}\check{k}} = 0\,.
\eea
Since 
\bea
0 = \delta_{\check{i}}{}^{\check{j}}{}_{,\check{k}} = \left(\gamma_{\check{i}\check{l}} \gamma^{\check{l}\check{j}}\right)_{,\check{k}} = \gamma_{\check{i}\check{l},\check{k}}\gamma^{\check{l}\check{j}}+\gamma_{\check{i}\check{l}}  \gamma^{\check{l}\check{j}}{}_{,\check{k}}\,,
\eea
Eq.~\eqref{eq:FirstDerivs} further yields 
\bea \label{eq:FirstDerivs2}
\big[\xi\big]:\,\,\gamma^{\check{m}\check{i}} \gamma_{\check{i}\check{l}} \gamma^{\check{l}\check{j}}{}_{,\check{k}} =\delta^{\check{m}}{}_{\check{l}} \gamma^{\check{l}\check{j}}{}_{,\check{k}} = \gamma^{\check{m}\check{j}}{}_{,\check{k}} =0\,. 
\eea
Furthermore, since Eqs.~\eqref{eq:FirstDerivs} and \eqref{eq:FirstDerivs2} are true everywhere along $\xi$, temporal derivatives can be added to yield
\bea \label{eq:FirstDerivs3}
\big[\xi\big]:\,\,\gamma_{\check{i}\check{j},\check{k}\check{t}} = 0 = \gamma^{\check{i}\check{j}}{}_{,\check{k}\check{t}}\,.
\eea
Equipped with these tools, we are now ready to tackle the first derivative terms in $G_{\check{t}\check{i}}$. Eq.~\eqref{eq:ConnectionExplicit} reduces covariant derivative to partial derivative, and then by Eqs.~\eqref{eq:FirstDerivs2} and \eqref{eq:FirstDerivs3}, we have
\begin{align}
&\big[\xi\big]:\,\,K_{\check{i}}{}^{\check{m}}{}_{|\check{m}}= -\frac{1}{2}\left(\gamma^{\check{m}\check{j}}\gamma_{\check{i}\check{j},\check{m} \check{t}}+ \gamma^{\check{m}\check{j}}{}_{,\check{m}}\gamma_{\check{i}\check{j},\check{t}}\right) = 0\,,\notag \\
&\big[\xi\big]:\,\, K_{|\check{i}}= -\frac{1}{2} \left(\gamma^{\check{j}\check{k}}\gamma_{\check{j}\check{k},\check{i} \check{t}}+ \gamma^{\check{j}\check{k}}{}_{,\check{i}}\gamma_{\check{j}\check{k},\check{t}}\right) = 0\,,
\end{align}
which are automatically regular without requiring any additional conditions. 

The spatial curvatures are then the only ones left, with the Ricci tensor given by 
\begin{align}\label{eq:RicciT}
\big[\xi\big]:\,\,{}^{(3)}R_{\check{j}\check{l}}={}^{(3)}R^{\check{i}}{}_{\check{j}\check{i}\check{l}} 
=& \sum_{\check{i}}\frac{{}^{(3)}R_{\check{i}\check{j}\check{i}\check{l}}}{ \gamma^e_{\iota \equalhat \check{i}}}
\,,
\end{align}
where the correspondence relation like $\iota \equalhat \check{i}$ means that the principal coordinate base $\partial_{\check{i}}$ should be along the eigenvector direction corresponding to $\gamma^e_{\iota}$. There is also a further contraction with the problematic $\gamma^{\check{i}\check{j}}$ to get to 
\begin{align}\label{eq:RicciS}
\big[\xi\big]:\,\,{}^{(3)}R= \sum_{\check{i}\check{j}}\frac{{}^{(3)}R_{\check{i}\check{j}\check{i}\check{j}}}{\gamma^e_{\iota \equalhat \check{i}}\gamma^e_{\iota' \equalhat \check{j}}} 
\,,
\end{align}
and subsequently ${}^{(3)}G_{\check{i}\check{j}}$. Because the three $\gamma^e_{\iota}$s generically decline at different rates, we need the Riemann tensor components in each term of the summations in Eqs.~\eqref{eq:RicciT} and \eqref{eq:RicciS} to separately decline sufficiently quickly. In fact, even if all the eigenvalues share the same rate of decline, there will still be $7$ regularity equations between Eqs.~\eqref{eq:RicciT} and \eqref{eq:RicciS}, but only $6$ independent components in the 3-D Riemann tensor, thus the equation set is over-determining, and generically only admit trivial solutions where each variable individually ``vanishes'' (sinks below the ``error budget''). In either case, we have explicitly
\begin{align} \label{eq:CondRiemann}
\big[\xi\big]&:\,\,{}^{(3)}R_{\check{i}\check{j}\check{i}\check{l}} = \mathcal{O}\left(\gamma^e_{\iota \equalhat \check{i}}\right)\,, \notag\\
\big[\xi\big]&:\,\,{}^{(3)}R_{\check{i}\check{j}\check{i}\check{j}} = \mathcal{O}\left(\gamma^e_{\iota \equalhat \check{i}}\gamma^e_{\iota' \equalhat \check{j}}\right)\,,
\end{align}
where $\check{i}$, $\check{j}$ and $\check{l}$ all take different values. Through index symmetries, Eq.~\eqref{eq:CondRiemann} accounts for all $6$ freedoms in the spatial Riemann tensor (explicitly, the $\check{i}=1,\,2$ or $3$ possibilities for the first line and the three inequivalent pairs $(\check{i},\check{j})=(1,2),\,(1,3)$ or $(2,3)$ for the second line).

These conditions can be further transcribed onto the second spatial derivatives of $\gamma_{\check{i}\check{j}}$. To this end, note that our principal coordinates are just rescalings of the Fermi coordinates, so the coordinate transformations between them is achieved via the Jacobian, and some simple Jacobian gymnastics allow us to import the standard Fermi result \cite{Ni:1978zz} to produce (note the sign difference with \cite{1963JMP.....4..735M}, stemming from the different conventions in the definition of the Riemann tensor)
\bea \label{eq:Offxi}
\gamma_{\check{i}\check{j}} = \Big[ \text{diag}_{\iota} \gamma^e_{\iota} \Big]_{\check{i}\check{j}}\Big|_{\xi} -\frac{1}{3}R_{\check{i}\check{l}\check{j}\check{m}}\Big|_{\xi}x^{\check{l}}x^{\check{m}} + \mathcal{O}\left((x)^3\right)\,,
\eea
that extend Eq.~\eqref{eq:SpecialSpatialCoord} off $\xi$ (the expansion coefficients labelled with $|_{\xi}$ are to be evaluated on $\xi$). Applying the Gauss-Codazzi equation, Eq.~\eqref{eq:Offxi} then implies 
\begin{align} \label{eq:SecondDerivAsRiemann}
\big[\xi\big]:\,\,\gamma_{{\check{i}\check{j},\check{p}\check{q}}} =& \frac{2}{3} R_{\check{i}(\check{p}\check{q})\check{j}} \\
=& \frac{2}{3}\left( {}^{(3)}R_{\check{i}(\check{p}\check{q})\check{j}}+K_{\check{i}(\check{p}}K_{\check{q})\check{j}}-K_{\check{i}\check{j}}K_{\check{p}\check{q}}\right)\,. \notag 
\end{align}
We note that there are four indices in the second derivatives of the metric, yet only three spatial dimensions to choose from, so at least one of the four indices repeat. On the other hand, if any index repeats three times or more, the 4-D Riemann tensor in the first line of the right hand side of Eq.~\eqref{eq:SecondDerivAsRiemann} vanishes due to its index antisymmetry properties. Applying these properties to the rest, and using the fact that the extrinsic curvature is diagonal on $\xi$, we obtain that all of the components in these second derivatives that are not automatically precisely zero are 
\begin{align} \label{eq:SecondDerivCons}
&\big[\xi\big]:\,\,\gamma_{{\check{i}\check{j},\check{i}\check{j}}} = \frac{1}{3} \left({}^{(3)}R_{\check{i}\check{j}\check{i}\check{j}}
+K_{\check{i}\check{i}}K_{\check{j}\check{j}}\right) 
= \mathcal{O}\left(\gamma^e_{\iota \equalhat \check{i}}\gamma^e_{\iota' \equalhat \check{j}}\right)
\,,  \notag \\
&\big[\xi\big]:\,\,\gamma_{{\check{i}\check{i},\check{j}\check{j}}} = -\frac{2}{3} \left({}^{(3)}R_{\check{i}\check{j}\check{i}\check{j}}
+K_{\check{i}\check{i}}K_{\check{j}\check{j}}\right) =\mathcal{O}\left(\gamma^e_{\iota \equalhat \check{i}}\gamma^e_{\iota' \equalhat \check{j}}\right)\,,  \notag \\
&\big[\xi\big]:\,\,\gamma_{{\check{i}\check{j},\check{i}\check{l}}} = \frac{1}{3} {}^{(3)}R_{\check{i}\check{j}\check{i}\check{l}}
=\mathcal{O}\left(\gamma^e_{\iota \equalhat \check{i}}\right)\,,  \notag \\
&\big[\xi\big]:\,\,\gamma_{{\check{i}\check{i},\check{j}\check{l}}} = -\frac{2}{3} {}^{(3)}R_{\check{i}\check{j}\check{i}\check{l}}=\mathcal{O}\left(\gamma^e_{\iota \equalhat \check{i}}\right)\,,  
\end{align}
where we have used Eq.~\eqref{eq:CondRiemann} and $\mathcal{C}_s^1$, and once again no two of $\check{i}$, $\check{j}$ and $\check{l}$ can equal each other. The conditions in Eq.~\eqref{eq:SecondDerivCons} can be summarized as

\vspace{3mm}
\noindent $\mathcal{C}^2_s$: \emph{
Spatial derivatives of the spatial metric vanish at least as quickly as the spatial metric itself as $\Sigma$ is approached, in the sense that, let $\check{p}$ be the doubly-repeated index appearing in the second spatial derivative of the spatial metric, then that second derivative must belong to $\mathcal{O}(\gamma^e_{\iota \equalhat \check{p}})$, and when there are two doubly-repeated indices (say $\check{p}$ and $\check{q}$), the derivative belongs to $\mathcal{O}(\gamma^e_{\iota \equalhat \check{p}}\gamma^e_{\iota' \equalhat \check{q}})$. } 
\vspace{3mm}
  
\noindent The first spatial derivatives of $\gamma_{\check{i}\check{j}}$ already vanish according to Eq.~\eqref{eq:FirstDerivs}, and these conditions on the second derivatives enforce a constraint on inhomogeneity in the early universe. 
Importantly, $\mathcal{C}_s^2$ is to be satisfied along every, and not just one, temporal coordinate curve of the Gaussian normal system. Just like $\beta_{,x}=0$ for some function $\beta(x)$ at one particular $x=x_0$ value would be a mere boundary condition that's not very constraining, having it satisfied everywhere will force $\beta$ to be a constant. In our case, there is a complication that the first derivatives are made to vanish due to the choice of the principal coordinate system, which is schematically akin to going into local coordinate patches $(x',y')_{q}$ individually rotated to adapt to the slope of $\beta$ ($x'$ axis is chosen to be parallel to this slope) at each point $q \in \beta$, so $\beta_{,x'}|_{q}=0$ is guaranteed whatever the shape of $\beta$ (besides being sufficiently smooth to allow derivatives). Now, the vanishing of the second derivative $\beta_{,x'x'}|_{q}=0$ carries the weight instead. It is a nontrivial condition that ensures the infinitesimally-close neighbouring local patches $(x',y')_{q\pm\delta q}$ do not need to be rotated against $(x',y')_{q}$ (Jacobian is identity). The same argument continues on and propagates out further away from $q$ if the vanishing of the second derivative is to be satisfied everywhere, so $\beta$ is forced to be a straight line again, that can be made into a constant if a boundary condition 
\bea \label{eq:ToyBC}
\beta_{,\hat{x}}|_{\hat{x}_0}=0
\eea
is supplied at any single point $\hat{x}_0$ in some global coordinate system $(\hat{x},\hat{y})$. 

This last step amounts to judiciously choosing the global/finite-regional coordinate system, which is necessary in our case also, since the metric is not spatially constant under arbitrary coordinate systems even for the FLRW spacetime. In particular, the metrics as they are written under polar coordinates in Eq.~\eqref{eq:ConfTrans} are spatially variable (the basis vectors for this coordinate system are not parallelly transported, thus there are many non-vanishing spin coefficients even in a flat spacetime), and cannot be directly plugged into $\mathcal{C}_s^2$ that is instead stated under the more physical principal coordinates (geodetically constructed, somewhat like Cartesian coordinates in flat spacetime). With the toy example, Eq.~\eqref{eq:ToyBC} can be achieved by simply extending the local $(x',y')_{q}$ for an arbitrary $q$ into a global coordinate system $(\hat{x},\hat{y})$, which in our context is mimicked by using the principal coordinates associated with an arbitrary $\xi$ within entire finite regions surrounding that geodesic.

When homogeneity is coupled with an initial ${K}>0$ (growing spatial volumes) allowed by $\mathcal{C}^1_s$, we have the basic ingredients underlying Wald's theorem \cite{1983PhRvD..28.2118W},
as a concrete realization of the more general cosmological ``no-hair'' conjecture \cite{1977PhRvD..15.2738G,1982PhLB..110...35H}, that shows isotropy and spatial flatness (local resemblance to de Sitter) can possibly be achieved later through accelerated expansion, due to inflation (with any vestiges plausibly manifesting as the low multipole temperature anomalies of the Cosmic Microwave Background \cite{2019arXiv190602552P}, provided those are not simply statistical fluctuations accentuated by cosmic variance). Within the proof of Wald's theorem, homogeneity is required to maintain ${}^{(3)}R \leq 0$, but can be slightly relaxed to allow small perturbations on top of a homogeneous background \cite{1983JETPL..37...66S,1983veu..conf..273B,1983veu..conf..267B,1988ASIC..219..261P}. 
Furthermore, if one is only interested in isotropy, then it has long been known \cite{1968ApJ...151..431M,1992PhR...214..223G} that anisotropy drops off rapidly with the effective spatial scale factor in a homogeneous universe, even without inflation. In this sense, isotropy may be seen as a secondary consequence of $\mathcal{C}^2_s$, provided that the universe subsequently expands.   

Finally we note that the conditions $\mathcal{C}_s^{1/2}$ are to be applied in conjunction with the generic condition $\mathcal{C}_g$. Namely that the spatial metric induced on $\Sigma$ from the two sides match up, and that the extrinsic curvature of the two sides should also suitably agree. For the latter condition, it is worth noting that, in principle, $\mathcal{C}_g$ allows a surface layer of radiation or gravitational impulsive wave \cite{1968AIHPA...8..327C,MTW,Penrose:1972ia,Penrose:1972xrn} to reside on the null surface $\Sigma$, permitting the extrinsic curvature to jump and the curvature tensors to become distributional at $\Sigma$ \cite{Taub:1980zr}. 
However, the specific condition $\mathcal{C}^{1}_s$ removes such scenarios. In other words, the discontinuities and mild ``zero-width blow-up'' \cite{Mikusinski1948} of a Dirac-delta type distribution become collateral casualties of our attempt to avoid more severe divergences. 

\subsection{The Klein-Gordon equation} \label{sec:CondKG}
\subsubsection{The conditions}
We have regularized the left hand side of the Einstein's equations in the last section, and now turn to the right hand side, the matter stress-energy. We also need to make sure that the equation of motion for the matter itself is well-behaved. As a tractable representative case (particularly relevant for those single field inflation scenarios without other fields before reheating), we concentrate on the scalar field, which satisfies the Klein-Gordon equation 
\bea
g^{\check{a}\check{b}}\varphi_{,\check{a}\check{b}}-g^{\check{a}\check{b}}\Gamma^{\check{c}}_{\check{a}\check{b}}\varphi_{,\check{c}}= \mathcal{V}'(\varphi)\,,
\eea
where the prime denotes derivative against $\varphi$. 
On $\xi$, the 4-D metric is block diagonal, so the equation becomes
\bea
\big[\xi\big]:\,\,
-\varphi_{,\check{t}\check{t}}+\gamma^{\check{i}\check{j}}\varphi_{,\check{i}\check{j}}+\Gamma^{\check{c}}_{\check{t}\check{t}}\varphi_{,\check{c}}-\gamma^{\check{i}\check{j}}\Gamma^{\check{c}}_{\check{i}\check{j}}\varphi_{,\check{c}}=\mathcal{V}'(\varphi)\,.
\eea
Since $\xi$ is a geodesic always at the origin of the $x^{\check{a}}$ coordinate system, and $\check{t}$ in this coordinate system is its affine parameter, we have by the geodesic equation that $\Gamma^{\check{c}}_{\check{t}\check{t}}=0$. Furthermore, from the same procedure that yielded Eq.~\eqref{eq:Offxi}, we see that, just as within the Fermi coordinates, the first spatial derivatives of $g_{\check{a}\check{b}}$ vanishes on $\xi$ (but different from the Fermi case, the temporal derivatives do not vanish, since our $\gamma_{\check{i}\check{j}}$ is not constant along $\xi$), so 
\bea \label{eq:ConnectionTime}
\big[\xi\big]:\,\,\Gamma^{\check{t}}_{\check{i}\check{j}} =\frac{1}{2}\gamma_{\check{i}\check{j},\check{t}}=-K_{\check{i}\check{j}}\,, \quad \Gamma^{\check{k}}_{\check{i}\check{j}} = 0\,.
\eea
Therefore, the Klein-Gordon equation reduces to 
\begin{align} \label{eq:KGDetails}
\big[\xi\big]:\,\,
\mathcal{V}'(\varphi)=& -\varphi_{,\check{t}\check{t}}+\gamma^{\check{i}\check{j}}\varphi_{,\check{i}\check{j}}-\frac{1}{2}\gamma^{\check{i}\check{j}}\gamma_{\check{i}\check{j},\check{t}}\varphi_{,\check{t}}
\notag \\
=& -\varphi_{,\check{t}\check{t}} + \sum_{\check{i}}\frac{\varphi_{,\check{i}\check{i}}}{\gamma^e_{\iota \equalhat \check{i}}}+K\varphi_{,\check{t}}
\,,
\end{align}
where the condition $\mathcal{C}^1_s$ ensures that the coefficient to the $\varphi_{,\check{t}}$ term is regular, so the temporal derivatives of $\varphi$ does not need to vanish. Only the spatial derivative needs to decline sufficiently quickly to ensure that the equation of motion admits a well-defined limit on $\Sigma$.

The other condition for $\varphi$ is that the stress-energy tensor ${T}_{\check{a}\check{b}}$ that equates to ${G}_{\check{a}\check{b}}$ in Eqs.~\eqref{eq:G1}-\eqref{eq:G3} should not diverge. Explicitly (including the contribution from a cosmological constant $\Lambda$)
\begin{align}\label{eq:EnergyKG}
\big[\xi\big]:\,\, T_{\check{t}\check{t}} =& \frac{1}{2}\varphi_{,\check{t}}^2 + \frac{1}{2}\gamma^{\check{i}\check{j}} \varphi_{,\check{i}} \varphi_{,\check{j}}+ \mathcal{V}(\varphi) + \frac{\Lambda}{8\pi}\,, \notag \\
\big[\xi\big]:\,\, {T}_{\check{i}\check{j}} =& \gamma_{\check{i}\check{j}}\left( \frac{1}{2}\varphi_{,\check{t}}^2 - \frac{1}{2}\gamma^{\check{k}\check{l}} \varphi_{,\check{k}} \varphi_{,\check{l}}- \mathcal{V}(\varphi) - \frac{\Lambda}{8\pi} \right) \notag \\
&+ \varphi_{,\check{i}}\varphi_{,\check{j}}\,, \notag \\
\big[\xi\big]:\,\, {T}_{\check{t}\check{i}} =& \varphi_{,\check{t}} \varphi_{,\check{i}}\,, 
\end{align}
and the only dangerous term is 
\bea \label{eq:KGCond2}
\big[\xi\big]:\,\,\gamma^{\check{k}\check{l}} \varphi_{,\check{k}} \varphi_{,\check{l}} = \sum_{\check{k}}\frac{(\varphi_{,\check{k}})^2}{\gamma^e_{\iota \equalhat \check{k}}}\,.
\eea
Combining with our earlier discussion on the Klein-Gordon equation, and noting that the $\gamma^e_{\iota}$s can decline at different rates (or that there are still two regularity equations arising from Eqs.~\ref{eq:KGDetails} and \ref{eq:KGCond2} even if they do share the same rate), but there is only one variable $\varphi$, we obtain the conditions

\vspace{3mm}
\noindent $\mathcal{C}^3_s$: \emph{The first and second spatial derivatives of $\varphi$ must vanish sufficiently quickly as compared to the spatial metric, in the sense that $\varphi_{,\check{i}} = \mathcal{O}\left((\gamma^e_{\iota\equalhat \check{i}})^{1/2}\right)$ and 
$\varphi_{,\check{i}\check{i}} = \mathcal{O}\left(\gamma^e_{\iota\equalhat \check{i}}\right)$.
} 
\vspace{3mm}

\noindent These spatial homogeneity conditions ensure that ${T}_{\check{i}\check{j}}$ and ${T}_{\check{t}\check{j}}$ vanish on $\Sigma$, but none of the terms in  
$T_{\check{t}\check{t}}$ need to. Unfortunately then, $\mathcal{C}_s^3$ alone is not sufficient to ensure potential energy dominance to launch inflation if $\varphi$ is the inflaton. 

\subsubsection{The inflationary universe} \label{sec:Inflation}
Nevertheless, additional supplementary junction conditions $\mathcal{C}_{\rm sup}$ can be obtained through physical considerations. Such conditions are not needed by the mathematical regularity of the various equations of motion, so not strictly the subject of the present paper. Nevertheless, they owe their appearance to $\mathcal{C}_s$, and are thus interesting to investigate. 

It is to be noted that the energy density $T_{\check{t}\check{t}}$ as given by Eq.~\eqref{eq:EnergyKG} is a special (scalar field) case of the matter density $\rho$ that appears in FLRW derivations (the FLRW comoving coordinates are Gaussian normal, so $\xi$ is automatically the worldline of a comoving observer), and will scale as $\rho\propto a^{-3(1+w)}$, when the equation of state is $P=w\rho$. So if $\rho$ is finite at $a=0$, it will vanish at later times unless $w \leq -1$. Looking at the same issue in reverse, if instead $\rho\neq 0$ when $a > 0$ with a $w > -1$, the FLRW will have a diverging Hubble's parameter $\mathcal{H}=a_{,t}/a$ on $\Sigma$, defying $\mathcal{C}_s^1$ which requires it to be regular. We can see this quite readily from the Friedmann equation 
\bea \label{eq:FRWMatch}
3\mathcal{H}^2 = 8\pi \rho - \frac{{}^{(3)}R}{2} + \Lambda\,.
\eea
Since we have required ${}^{(3)}R$ to remain regular on $\Sigma$ (for FLRW, this requires flatness $\kappa=0$, thus ${}^{(3)}R|_{\Sigma}=0$), and $\Lambda$ is just a constant, there is nothing to cancel with the divergence from $\rho$. Furthermore, adding anisotropy would unlikely be helpful, because while it adds a shear scalar term \cite{1983PhRvD..28.2118W}
\begin{align}
\sigma^2 \equiv& \frac{1}{2}\left( K_{\check{i}\check{j}}-\frac{1}{3}K\gamma_{\check{i}\check{j}} \right) \left( K_{\check{k}\check{l}}-\frac{1}{3}K\gamma_{\check{k}\check{l}} \right) \gamma^{\check{i}\check{k}}\gamma^{\check{j}\check{l}} \notag \\
=& \frac{1}{2}\text{tr}(K\cdot K)-\frac{1}{6}K^2
\end{align}
into the right hand side of Eq.~\eqref{eq:FRWMatch}, this term is regulated by $\mathcal{C}_s^1$ to be non-divergent. 

In summary, $\mathcal{C}_s^1$ and the fact that our universe has a non-vanishing matter energy density today, together, force a scalar inflaton field $\varphi$ to be the only matter near $\Sigma$ (c.f.,~Ref.~\cite{2002PhRvD..65h3507A}), which must also behave like a perfect cosmological constant with $w=-1$ (we ignore the $w <-1$ case since there are no accepted matter models with that kind of equation of state). 
This makes physical sense, since traditional particles of constant finite spatial metric sizes (the standard assumption is that the sizes of particles like electrons are determined by local physics and will not scale with the cosmic size $a$) shouldn't already exist all the way back at $\Sigma$, or else they will each engulf the entire spatial slice and overlap with one another, at the very least significantly deviate from our normal intuition of how they behave. A potential-energy-dominated inflaton field or cosmological constant do not need to possess any finite-spatial-size features on the other hand, and can be accommodated quite easily. They will also not dilute or concentrate, so won't produce diverging stress-energy tensors when $a=0$. We therefore impose the condition 

\vspace{3mm}
\noindent $\mathcal{C}_{\rm sup}: $\emph{
$\varphi_{,\check{a}} \rightarrow 0$ sufficiently quickly so $T_{\check{t}\check{t}}|_{\Sigma}$ as given by Eq.~\eqref{eq:EnergyKG} is contributed only by the potential $\mathcal{V}$ and the cosmological constant. }
\vspace{3mm}

\noindent 
This condition translates directly into the Cartesian coordinates for FLRW (recall $\kappa=0$), which coincide with the principal coordinates associated with the timelike geodesic at the arbitrarily chosen spatial origin. As a consequence, the Lorentzian universe will be born directly into an inflationary period with (solving Eq.~\ref{eq:FRWMatch} for the FLRW case)
\bea \label{eq:Infl}
a(t) \approx \mathcal{B} e^{\lambda t}, \quad \lambda=\mathcal{H}=\sqrt{\frac{8\pi \mathcal{V}|_{\Sigma}+\Lambda}{3}}\,,
\eea
where $\mathcal{B}$ is a constant, and the approximation is valid regardless of the shape of the potential $\mathcal{V}$, because $\mathcal{C}_{\rm sup}$ sets inflation off with an instantaneous no-rolling configuration $\varphi_{,t}|_{\Sigma}=0$. However, since (c.f.~Eq.~\ref{eq:KGDetails})
\bea \label{eq:KGFLRW}
\varphi_{,tt}=-3\mathcal{H}\varphi_{,t}-\mathcal{V}'
\eea
does not need to vanish initially, the $\varphi$ field will eventually begin to roll. A flattish $\mathcal{V}$ could significantly prolong inflation though, depending on the location of the flat region in relation to $\varphi|_{\Sigma}$, either by reducing initial $\varphi_{,tt}|_{\Sigma}=-\mathcal{V}'|_{\Sigma}$ to delay rolling (if $\mathcal{V}'|_{\Sigma} =0$ precisely, the Klein-Gordon Eq.~\ref{eq:KGFLRW} is satisfied at all times without $\varphi$ ever changing), or/and to allow the $\varphi$ field to settle into a standard slow-roll regime of $\varphi_{,t}\approx -\mathcal{V}'/(3\mathcal{H})$ at a later time. Regardless, the constraints on regularity within route B, through $\mathcal{C}_{\text{sup}}$ (as a consequence of $\mathcal{C}_s^1$) specifically, compels inflation to start without delay (in reverse, such an inflationary homogeneous early universe is in compliance with all the conditions in this paper). 
I.e., there isn't a pre-inflationary radiation- or kinetic-dominated deceleration phase, the signatures of which had been searched for, but indeed not found in observational data \cite{2018arXiv180706211P}. Furthermore, the conditions $\mathcal{C}^{2}_{s}$, $\mathcal{C}^{3}_{s}$ and $\mathcal{C}_{\text{sup}}$ are beneficial to the inflation paradigm in another sense, that they could conceivably take us to the required initial homogeneity \cite{1992PhR...214..223G,2000PhRvD..61b3502V,2003PhRvD..67h3515A} (a more precise quantitative and non-perturbative statement of this requirement would facilitate further analysis).

\section{Discussion and conclusion} \label{sec:Conclusion}
In this paper, we have examined what classical junction conditions would be required for a transition of our universe into a purely timelike Riemannian regime through the big bang. 
So far, the restrictions they impose do not appear to raise immediate contradictions that would spoil the viability of the signature change scenario in terms of describing our physical universe. Instead, useful constraints seem to arise. E.g., the conditions $\mathcal{C}_s^1$ and $\mathcal{C}_s^2$ (particularly in the form of Eq.~\ref{eq:CondRiemann}) enforce that as we approach $\Sigma$ along a timelike geodesic $\xi$, the geodetically developed spatial slices in the principal coordinate system associated with $\xi$ become intrinsically and extrinsically flat. Because the direction of $\xi$ can be chosen freely ($\xi$ is any temporal coordinate curve of any Gaussian normal system, which can be built out of arbitrary timelike congruences), this means that, via the Gauss-Codazzi equation, the projection of the 4-D covariant Riemann curvature tensor onto any spatial tangent plane at a point near $\Sigma$ must be small (when written in sensible coordinates whose Jacobian against the principal coordinates associated with the timelike geodesic orthogonal to that plane does not diverge). Although the spatial projection operator is rank deficient, its kernel is only one dimensional (specifically the tangential direction to $\xi$; the projection will yield zero for nonvanishing vectors only if the vector is precisely along this direction), so if the projection is vanishingly small for any arbitrary $\xi$, the full 4-D covariant Riemann tensor should be nearly zero (because any large component hidden inside the kernel of one projection operator would have been exposed by a different operator). 
In this sense, a strong version of the low gravitational entropy condition for the early universe, mentioned in item $3$ of Sec.~\ref{sec:Intro2}, is realized. In particular, the inflationary FLRW discussed in Sec.~\ref{sec:Inflation}, that's compatible with $\mathcal{C}_s^{1/2}$, not only has a vanishing 4-D Weyl curvature as FLRW metrics always do due to their symmetries, but the entire 4-D Riemann curvature vanishes when $a\rightarrow 0$. In contrast, this is not the case with dust or radiation dominated FLRWs that do not satisfy $\mathcal{C}_s^{1}$. 

Our conditions $\mathcal{C}^{1/2/3}_s$ and $\mathcal{C}_{\rm sup}$ are however not yet as strong as they can be. While they ensure the existence of one-sided limits such as ${}^{\mp}G_{ab}|_{\Sigma}$, so that the equations of motion for metric and matter can be extended onto the big bang $\Sigma$ from either side, they do not require that the ${}^{\mp}$ limits match up, which would force the two signature regimes to connect up in a smoother manner. 
This omission is intentional (besides trying to be conservative given our ignorance of whether the matching is absolutely necessary), because then the one-sided conditions enumerated in this paper would admit physical interpretations independent of signature change. 
Namely, they need to be satisfied if the equations of motion are to be extended onto the big bang itself. Without including the big bang into the domain of validity for these equations, the Lorentzian universe will become an open set, without a suitable boundary to impose boundary (initial) conditions on. In other words, regardless of one's view on what happens beyond the big bang, the main result of this paper can be read as necessary conditions for our Lorentzian universe to admit a Cauchy description. The utility of this paper thus does not fully diminish even if the signature change scenario is not physically realized in nature. 

There are many important issues that we have not been able to tackle. In particular, unlike in \cite{1992CQGra...9.1535E,1992GReGr..24.1047E}, where the genuinely classical transition into a spacelike Riemannian region occurs prior to the Planck time, it seems more difficult for us to circumvent the issue of quantum gravity, because the scale factor do need to vanish in our case. The theory of quantum gravity is as yet unavailable, thus our discussion merely aims to shed some light on the possible behaviour of the classical saddle point solutions that hopefully would dominate the full quantum path integral. Having said that, it must be noted though, that whether such a semi-classical approach even makes sense in the gravitational context is presently subject to debate \cite{2017PhRvL.119q1301F,2017PhRvD..96d3505D}. 
Furthermore, one could also note that the criteria for the onset of quantum gravity, based on dimensional analysis, is not Lorentz invariant unless one demands macroscopically separated events connected by null rays also be treated quantum gravitationally \cite{2002GReGr..34.2043H}, a prospect that has not been shown to be necessary. Taken to the extreme, this appears to indicate that the distances computed with the Lorentzian metric may not be the sole determining factor regarding the onset of quantum gravity, and one should perhaps be more circumspect when stating that quantum gravity must be evoked near the transition surface, which in our case could just be another macroscopic null surface. In other words, the trans-Plankian problem \cite{2001PhRvD..63l3501M} of inflation might not necessarily arise. 

Finally, even staying at the purely classical level, the junction conditions examined in this paper are minimal, in that while they ensure initial conditions can be imposed on the big bang, they do not tell us whether the evolution off of such compliant (with the junction conditions) initial data sets can be a well-posed initial value problem. In other words, they do not guarantee that physically interesting solutions (not plagued by wild exponentially growing perturbations, which inevitably lead to an extreme prevalence of singularities that appear to arise spontaneously) exist (the inflationary FLRW do satisfy the junction conditions, but its stability may need further scrutiny within our context). 
For different purposes, the required level of well-posedness is different. When trying to simulate the universe on a computer, initial conditions even off of the constraint surface (i.e., do not strictly satisfy the Hamiltonian and momentum constraints) are relevant, since numerical errors are inevitable, not least because computers cannot store numbers to infinite digits (i.e., we always have truncation error). There is of course the possibility that our physical universe is not amenable to being studied this way, and the well-posedness condition can presumably be relaxed to considerations on only a neighbourhood of the constraint-satisfying initial conditions space, surrounding that of our actual universe. Regardless, answering this well-posedness question demands substantial technical dexterity (as attested by the already strenuous work that went into proving the well-posedness of specific formulations of Einstein's equations off more familiar spacelike Cauchy surfaces), and will have to be addressed in future works. 

\acknowledgements
This work is supported by the National Natural Science Foundation of China grants 11503003 and 11633001, the Interdiscipline Research Funds of Beijing Normal University, and the Strategic Priority Research Program of the Chinese Academy of Sciences Grant No. XDB23000000. 


\bibliography{paper.bbl}




\end{document}